\documentclass[useAMS,usenatbib]{mn2e}
\usepackage{epsf}
\usepackage{amsmath}
\usepackage{subfigure}

\newcommand{\hmpc}{h^{-1}{\rm Mpc}}

\newcommand{\kms}{\;{\rm km}\,{\rm s}^{-1}}

\newcommand{\msun}{M_{\odot}}

\newcommand{\lcii}{l_{\cii}}
\newcommand{\lciv}{l_{\civ}}

\newcommand{\aap}{A\&A}
\newcommand{\apjs}{ApJS}
\newcommand{\apj}{ApJ}
\newcommand{\apjl}{ApJL}
\newcommand{\aj}{AJ}

\newcommand{\mnras}{MNRAS}

\newcommand{\pasp}{PASP}

\newcommand{\hi}{\mathrm{H\,I}}

\newcommand{\cii}{\mathrm{C\,II}}
\newcommand{\ciii}{\mathrm{C\,III}}
\newcommand{\civ}{\mathrm{C\,IV}}
\newcommand{\siiv}{\mathrm{Si\,IV}}
\newcommand{\cv}{\mathrm{C\,V}}

\newcommand{\ocii}{\Omega_{\mathrm{C\,II}}}
\newcommand{\ociv}{\Omega_{\mathrm{C\,IV}}}
\newcommand{\fcii}{f_{\mathrm{C\,II}}}
\newcommand{\fciv}{f_{\mathrm{C\,IV}}}
\newcommand{\lognciv}{\log(N_{\mathrm{C\,IV}})}

\newcommand{\fesc}{f_{\mathrm{esc}}}
\newcommand{\fescfive}{f_{\mathrm{esc,5}}}

\newcommand{\heii}{\mathrm{He\,II}}

\newcommand{\tes}{\tau_{\mathrm{es}}}

\newcommand{\xhiv}{x_{\mathrm{H\,I,V}}}

\begin{document}

\title[Carbon Reionization]{The Reionization of Carbon}

\author[K.\ Finlator et al.]{
\parbox[t]{\textwidth}{\vspace{-1cm}
Kristian Finlator$^{1,8}$, 
Robert Thompson$^{5}$,
Shuiyao Huang$^{2}$,
Romeel Dav\'e$^{5,6,7}$,
E.\ Zackrisson$^3$,
B.\ D.\ Oppenheimer$^{4}$
}
\\\\$^1$ Dark Institute of Cosmology, Niels Bohr Institute, University of Copenhagen, Copenhagen, Denmark
\\$^2$ University of Massachussetts, Amherst, MA, USA
\\$^3$ The Oskar Klein Centre, Department of Astronomy, AlbaNova, Stockholm University, SE-106 91 Stockholm, Sweden
\\$^4$ CASA, Department of Astrophysical and Planetary Sciences, University of Colorado, 389-UCB, Boulder, CO 80309, USA
\\$^5$ University of the Western Cape, Bellville, Cape Town 7535, South Africa
\\$^6$ South African Astronomical Observatories, Observatory, Cape Town 7525, South Africa
\\$^7$ African Institute for Mathematical Sciences, Muizenberg, Cape Town 7545, South Africa
\\$^{8}$ kfinlator@dark-cosmology.dk
\author[CII and CIV at $z\geq5$]{
K.\ Finlator,
R.\ Thompson,
S.\ Huang,
R.\ Dav\'e,
E.\ Zackrisson,
\& B.\ D.\ Oppenheimer
}
}

\maketitle

\begin{abstract}
Observations suggest that $\cii$ was more abundant than $\civ$ in the 
intergalactic medium towards the end of the hydrogen reionization epoch 
($z\sim6$).  This transition provides a unique opportunity to study the 
enrichment history of intergalactic gas and the growth of the ionizing 
background (UVB) at early times.  We study how carbon absorption evolves 
from $z=10$--5 using a cosmological hydrodynamic simulation that includes 
a self-consistent multifrequency UVB as well as a well-constrained model 
for galactic outflows to disperse metals.  Our
predicted UVB is within $\sim2$--$4\times$ of that 
from~\citet{haa12}, which is fair agreement given the uncertainties.  
Nonetheless, we use a calibration in post-processing to account for
Lyman-$\alpha$ forest measurements while preserving the 
predicted spectral slope and inhomogeneity.  The
UVB fluctuates spatially in such a way that it always exceeds the 
volume average in regions where metals are found.  This implies both that 
a spatially-uniform UVB is a poor approximation and that metal absorption 
is not sensitive to the epoch when HII regions overlap globally even at 
column densites of $10^{12}$ cm$^{-2}$.  We find, consistent with 
observations, that the $\cii$ 
mass fraction drops to low redshift while $\civ$ rises owing the combined
effects of a growing UVB and continued addition of carbon in low-density 
regions.  This is mimicked in absorption statistics, which broadly agree 
with observations at $z=6$--3 while predicting that the absorber column 
density distributions rise steeply to the lowest observable columns.
Our model reproduces the large observed scatter 
in the number of low-ionization absorbers per sightline, implying that 
the scatter does not indicate a partially-neutral Universe at $z\sim6$.
\end{abstract}


\section{Introduction} \label{sec:intro}
Intergalactic $\civ$ and $\cii$ absorbers are observed along sightlines to 
quasars out to the highest redshifts probed~\citep{son01,ry06,bec09,rya09,sim11,dod13,bec11}.  
They occur naturally in cosmological hydrodynamic simulations in which
galactic outflows expel metals out to the virial radius and 
beyond~\citep{the02,cen05,opp06,opp08,opp09,cen11}, opening up 
the possibility of using models to interpret them as tracers of star 
formation, outflows, and the ultraviolet ionizing background (UVB).  
$\civ$ is the easier ion to observe owing to its larger oscillator
strength and greater redshift separation from the Lyman-$\alpha$ forest.  
It is also relatively straightforward to model because it arises in gas that 
is optically thin to ionizing photons and lies far enough away from 
galaxies that the local radiation field may be neglected to first order.  
Its overall mass density is robustly observed to decline slowly with 
increasing redshift~\citep{son01,bec09,rya09,dod10,dod13,sim11,bec11,coo13}. 
Numerous studies have shown that the observed decline can readily 
be accommodated by numerical simulations~\citep{opp06,opp09,cen11}.  There
remains some controversy as to what causes it: Some models indicate a 
combination of declining metallicity and increasing 
density~\citep{opp06,opp09} while others prefer an evolving mix of 
photoionization and shock-heating~\citep{cen11}.  Future high-resolution 
observations of $\civ$ absorbers past $z=3$ will constrain their velocity 
widths, which in turn can distinguish between different ionization 
mechanisms.

$\cii$ has received less theoretical and observational attention up until 
now.  This is due to change as observations push into the hydrogen reionization 
epoch, where physical conditions favor increasingly neutral ionization states.  
For example,~~\citet{bec11} have recently constrained the mass density of $\cii$, 
$\ocii$, to be $\geq9\times10^{-9}$ over the range $z=$5.3--6.4.  For 
comparison,~\citet{dod13} find that $\ociv$ is $8.4\times10^{-9}$ over the
interval $z=$5.30--6.20.  Hence by $z=6$, $\civ$ may already be subdominant 
to $\cii$.
Note that correcting these measurements for incompleteness would strengthen
the result:~\citet{bec11} estimate that they are $>50\%$ complete for $\cii$
absorbers with column densities above $10^{13.5}$cm$^{-2}$, 
whereas~\citet{dod13} estimate 70\% completeness for $\civ$ absorbers 
stronger than $10^{13.4}$cm$^{-2}$.  We will show (Figure~\ref{fig:Cfracs})
that $\cii$ is expected to become increasingly dominant to higher redshifts.
Hence tracing the IGM metallicity into the reionization epoch essentially
requires observations to shift from high- to low-ionization ions.

There is an additional reason to consider $\cii$ absorbers: At high 
redshift, $\cii$ traces lower-mass galaxies than $\civ$ because only 
more massive galaxies can expel metals out to the low densities where 
$\civ$ is the dominant ionization state; gas expelled by lower-mass 
galaxies remains at higher densities that favor $\cii$.  In 
particular,~\citet{opp09} (hereafter ODF09) find
that $\cii$ absorbers at $z=6$ are associated with galaxies with stellar 
masses of $\sim10^7\msun$ whereas $\civ$ is seen near galaxies with
$M_*\sim10^8\msun$ (their Figures 21 and 23).   The latter prediction
is supported by observations that the two strongest $\civ$ absorbers at
$z\sim6$ are physically associated with Lyman-$\alpha$ emitters~\citep{dia14}, 
whose stellar mass is characteristically $10^8$--$10^{10}\msun$~\citep{lid12}.  
The tendency for low-ionization metal absorbers to trace predominantly 
lower-mass galaxies~\citep{bec11,fin13,kul13}
immediately promotes them to a critical constraint on the origin of the 
UVB and of cosmological reionization because current models for 
reionization tend to rely heavily on low-mass galaxies to provide
the required flux of ionizing 
photons~\citep{fin11,alv12,kuh12,jaa12,rob13,duf14,wis14}.  Such 
systems are generally too faint to be observed in emission, hence 
low-ionization absorbers will remain the only way to constrain their 
abundance for the foreseeable future.

These considerations motivate a closer look at the ionization state of
carbon past $z=5$.  How does the fraction of intergalactic carbon that
is in $\cii$ and $\civ$ evolve in time?  How does this evolution map into
absorption statistics? Does it contain the signature of reionization, or 
is the ionization state of carbon primarily determined by local sources?  
How many additional absorbers will be identified in future observations 
that probe to fainter column densities?

Modelling metal absorbers in the reionization epoch requires a treatment
for the impact of local sources because the mean free path for ionizing
photons drops below 10 proper Mpc for $z>5$~\citep{wor14}. ODF09
explored the impact of a local field in a hydrodynamic simulation by 
assuming that each gas parcel's ionization state is dominated by the 
closest galaxy's radiation field (their \emph{Bubble} model).
They found that the local radiation field was much harder than the 
uniform~\citet{haa01} UVB, leading broadly to more highly-ionized 
gas in which the $\cii$ and $\civ$ column density distributions (CDDs) 
were respectively steeper and shallower.  In particular, the predicted
$\civ$ CDD was found to vary with column $N_{\civ}$ as $N_{\civ}^{\alpha}$
with $\alpha = -1.4$ and -2.0 for their local field model versus 
the~\citet{haa01} model.  This result is quite intuitive: accounting 
for the fact that galaxies can amplify the UVB locally boosts the 
ionization parameter in outflowing gas, increasing the number of 
strong high-ionization systems at the expense of the low-ionization 
systems.  Recently, ~\citet{dod13} measured the slope of the $\civ$ CDD
at $z\sim6$ and found $\alpha=-1.7\pm0.2$, favoring a somewhat flatter 
CDD (though not as flat as the \emph{Bubble} model) and indicating a 
significant ionizing contribution from local sources.  This model was 
therefore clearly a step in the right direction.

However, it neglected two important effects.  First, galaxies are 
clustered such that, by $z=6$, an overdense gas parcel will be irradiated 
by many nearby galaxies rather than just the nearest 
one~\citep{bar04,fur04a,fur04b,fur05}.  Second, gas
that resides near galaxies is often dense enough to attenuate the weak 
reionization-epoch UVB before it reaches carbon atoms, boosting the 
predicted abundance of $\cii$.  Working out which of these effects 
dominates requires improved theoretical models.

Recently, numerical simulations have acquired the ability to resolve the 
Jeans scale at the relevant redshifts while modeling the 
spatially-inhomogeneous UVB owing to stars and active galactic nuclei 
(AGN) self-consistently.  These improvements remedy both of the problems 
mentioned above.  In this paper, we use an updated model for the 
growth of structure and the UVB to study the ionization state of 
intergalactic carbon at $z\geq5$.
The outline of this work is as follows: In Section~\ref{sec:sims}, we 
summarize our numerical model.  In Section~\ref{sec:J}, we discuss our 
simulated UVB, compare it to the spatially-homogeneous~\citet{haa12} 
model (hereafter HM12), and explain how we use the latter to calibrate our 
UVB.  In Section~\ref{sec:compHM12}, we compare the predicted $\cii$ and 
$\civ$ fractions using our simulated UVB versus the HM12 UVB.  In 
Section~\ref{sec:obs}, we compare the predicted and observed absorber 
abundances.  In Section~\ref{sec:predict}, we then use the simulations 
to predict how the abundances of $\cii$ and $\civ$ absorbers evolve in
time.  We also study the scatter in the number of absorbers per line of
sight and produce predictions for the number of systems per survey.  
Finally, we summarize in Section~\ref{sec:sum}.

\section{Simulations}\label{sec:sims}
\subsection{Hydrodynamics, Star Formation, and Feedback}\label{ssec:hydro}
Our simulation was run using a custom version of 
{\sc Gadget-3} (last described in~\citealt{spr05}).  
Hydrodynamics is modelled using a density-independent formulation of 
smoothed particle hydrodynamics (SPH) that treats fluid instabilities 
accurately~\citep{hop13}.  We compute the physical properties of each
gas particle using a 5th-order B-spline kernel that incorporates information
from up to 128 neighbours.  Gas particles cool 
radiatively owing to collisional excitation of hydrogen and helium using 
the processes and rates in Table 1 of~\citet{kat96}, except that we 
relax the assumption of ionization equilibrium.  We model metal-line 
cooling using the collisional ionization equilibrium tables 
of~\citet{sut93}.

Gas whose proper hydrogen number density exceeds 0.13 cm$^{-3}$ acquires
a subgrid multiphase structure~\citep{spr03} and forms stars at a rate 
that is calibrated to match the observed Kennicutt-Schmidt law.  We 
model metal enrichment owing to supernovae of Types II and Ia as well 
as asymptotic giant branch stars; see~\citet{opp08} for details.
Galactic outflows form via a monte carlo model in which star-forming 
gas particles receive ``kicks" in momentum space and are thereafter 
temporarily decoupled hydrodynamically.  The outflow rates and 
velocities follow the ``ezw" prescription introduced in~\citet{dav13}.  

Our simulation discretizes the matter in a $6\hmpc$ volume using
$2\times256^3$ particles, which resolves the hydrogen-cooling limit at 
$10^8\msun$ with 65 particles.  We generate the initial conditions 
using an~\citet{eis99} power spectrum at $z=249$.  We run the simulation
to $z=3$, but we will focus mainly on results from snapshots taken at
redshifts between 5 and 10.  Our adopted cosmology is one in which 
$\Omega_M=0.3$, $\Omega_\Lambda=0.7$, $\Omega_b = 0.045$, 
$h=0.7$, $\sigma_8 = 0.8$, and the index of the primordial power spectrum 
$n=0.96$.

\subsection{Radiation Transport}\label{ssec:rt}
We discretize the radiation field owing to galaxies spatially on a regular 
grid of $32^3$ voxels and spectrally into 16 frequency groups spaced evenly 
between 1--10 Ryd.  Its evolution is followed via a moment method that 
couples with the opacity field from the gas in a way that captures the 
UVB's feedback effect~\citep{fin11}.  The 
emissivity of star-forming gas particles is proportional to their 
star formation rate, with a metallicity dependence that is computed from a 
modified version of {\sc Yggdrasil}~\citep{zac11}.  The emissivity is tabulated 
at 7 distinct metallicities between $Z=0$--$0.04$, and each gas particle's 
actual emissivity is computed by interpolating to its $Z$.  The $Z=0$ 
emissivity comes from~\citet{sch02};
the $Z=10^{-5}$ and $Z=10^{-7}$ emissivities come from~\citet{rai10}; and 
the $Z=0.001$--$0.040$ emissivities come from Starburst 99~\citep{lei99}, 
running with the Geneva tracks (high mass-loss version, without rotation) 
and Pauldrach/Hillier atmospheres.  Each model is adjusted to a Kroupa IMF.  

The most uncertain aspect of our model for the radiation field is the
ionizing escape fraction from galaxies, $\fesc$.  Choosing a constant value 
for all masses and redshifts leads to a reionization history that either 
begins too late or overproduces the observed UVB amplitude at 
$z<6$~\citep{fin11,kuh12}.  In order to match both constraints, recent 
models assume that the volume-averaged $\fesc$ varies in 
time~\citep{kuh12,haa12,mit13}; this behavior is seen directly in some 
models~\citep{so14}.  Alternatively, a number of detailed theoretical models 
suggest that $\fesc$ is larger at lower 
masses~\citep{wis09,yaj11,wis14,paa13,fer13}.  Given
that the typical mass of star-forming halos increases with time, this leads
naturally to a scenario in which the mean $\fesc$ decreases, yielding
good agreement with observations~\citep{alv12}.

We have found in practice that a pure mass dependence is not sufficient
to match available constraints owing to the strong suppression of star
formation by galactic outflows at low masses, so we adopt a
model for $\fesc$ that includes both mass- and redshift-dependence.  We
divide halos into three mass ranges.  \emph{Minihalos} ($M_h \leq 10^8 \msun$)
have $\fesc = 0.8$ at all times. \emph{Photoresistant} halos 
($M_h \geq 10^{9.5}\msun$) have an $\fesc$ that varies with redshift:
\begin{eqnarray} \label{eqn:fesc}
\fesc = \left\{
\begin{array}{lc}
  0.5 & z \geq 10 \\
  \fescfive\left(\frac{1+z}{6}\right)^\kappa & z < 10
\end{array} \right.
\end{eqnarray}
Here, the $\fesc$ from photoresistant halos at $z=5$ is 
$\fescfive=0.05$ and the slope of the redshift dependence is 
$\kappa=3.8$.  The $\fesc$ from \emph{photosensitive} halos 
($10^8 < M_h/\msun < 10^{9.5}$) varies linearly with $\log(M_h/\msun)$
between 0.8 at $10^8\msun$ and the redshift-dependent value
for photoresistant halos.  This model associates 
high $\fesc$ values with lower masses and earlier times.  

Note that, in our model, $\fesc$ does not depend on frequency.  This contrasts
with HM12, who assume 
that no photons with energies above 4 Ryd escape from galaxies.  Our
choice is motivated by theoretical models in which ionizing radiation
escapes either by leaking through porous ISMs or because some stars
have irregular orbits that take them outside the ISM's dense
regions~\citep{wis09}.  In both scenarios, the wavelength dependence
of $\fesc$ could be weak.  In practice, however, we have verified that 
this assumption has negligible impact on our predictions regarding 
$\civ$ and $\cii$ absorbers at $z\sim6$, hence we will not discuss 
it further.

We model the AGN radiation field via a spatially-averaged
radiation transfer calculation (see equation 1 of HM12).  The field
is discretized spectrally into the same frequency bins as the 
galaxy field but is spatially uniform (except in dense regions, where
it is attenuated by a self-shielding model that we describe below).  
The emissivity and spectral shape are taken 
from Equations 37 and 38 of HM12, with the modification that the
emissivity vanishes at $z>8$ (this assumption does not matter because 
galaxies dominate the UVB at energies less than 10 Ryd for $z>4$).  The 
sink term is given by the simulation's volume-averaged opacity.
The transfer calculation is included in the simulation's cooling/ionization 
iteration, hence it is fully coupled into the radiation hydrodynamic framework.

The radiation field is not modelled with high enough spatial resolution to
capture self-shielding in Lyman limit systems, hence we attenuate it in 
dense regions using a 
subgrid prescription based on the local Jeans length that is a 
generalization of the model presented in~\citet{sch01}.  This model is not 
important for $\civ$ absorbers as they arise primarily in the circumgalactic 
medium.  On the other hand, it is crucial for low-ionization 
absorbers~\citep[Figure 1 of][]{fin13}, which arise in denser gas.

The optical depth $\tau_\nu$ through a region in hydrostatic equilibrium
with Jeans length $L_\mathrm{J}$ owing to a set of species $n_i$ with 
absorption cross sections $\sigma_{\nu,i}$ is given by
\begin{eqnarray}\label{eqn:tau1}
\tau_\nu = L_\mathrm{J} \sum_i n_i \sigma_{\nu,i},
\end{eqnarray}
where we sum only over the abundances of neutral hydrogen and neutral and 
singly-ionized helium.  To compute the local ion abundances $n_i$, we 
assume that gas is in photoionization equilibrium (note that this 
assumption is only used to model self-shielding; the actual ionization 
states of hydrogen and helium are tracked using our nonequilibrium 
ionization solver).  Following HM12, we define the 
following dimensionless ionization parameters:
\begin{eqnarray*}
R_\mathrm{HI} & = & \frac{\Gamma_\mathrm{HI}}{n_e \alpha_\mathrm{HII}} \\
R_\mathrm{HeI} & = & \frac{\Gamma_\mathrm{HeI}}{n_e \alpha_\mathrm{HeII}} \\
R_\mathrm{HeII} & = & \frac{\Gamma_\mathrm{HeII}}{n_e \alpha_\mathrm{HeIII}}
\end{eqnarray*}
With these definitions, Equation~\ref{eqn:tau1} may be expanded to
\begin{eqnarray}\label{eqn:tau2}
  \begin{aligned}
    \tau_\nu = & L_\mathrm{J} n_H \times \\
               & \left[\frac{\sigma_{\nu,HI}}{(1 + R_\mathrm{HI})} + \frac{1-X_H}{4 X_H} \frac{(\sigma_{\nu,HeII} + R_\mathrm{HeI}\sigma_{\nu,HeIII})}{(1 + R_\mathrm{HeI} + R_\mathrm{HeI}R_\mathrm{HeII})} \right].
  \end{aligned}
\end{eqnarray}
In order to evaluate equation~\ref{eqn:tau2}, we need
the photoionization rates $\Gamma_i$, gas temperature, and
electron abundance.  We adopt the local values for each particle,
all of which are computed self-consistently by the simulation.  In 
Appendix~\ref{app:ss}, we show that this treatment is in good agreement
with the spatially-resolved radiation transport calculations 
of~\citet{rah13}.

Although our simulation volume is too small to be numerically resolved, 
it yields a plausible reionization history.  The bottom panel of 
Figure~\ref{fig:Jamp} shows the time-evolution of the volume-averaged 
neutral fraction $\xhiv(z)$.  The predicted reionization history is 
extended, with $\xhiv$ dropping below 90\% at $z\sim10$ and 50\% shortly 
after $z=8$.  The predicted optical depth to Thomson scattering $\tes$
is 0.057, which is barely within the observed $2\sigma$ confidence
interval of $0.081\pm0.012$~\citep{hin13}.  This value is considerably lower
than our previous simulation, which yielded 0.071~\citep{fin13}.  The 
difference owes to varying degrees to the lower adopted values for $\sigma_8$
and $\Omega_b$ (0.08 and 0.045 versus 0.082 and 0.046, respectively); our
smaller cosmological volume (our current simulation volume subtends
6 versus 9$\hmpc$); and our different model for the ionizing escape 
fraction, which is capped at 0.8 rather than 1.0 and decreases to high 
masses rather than being constant at all masses.  Additionally, the ezw 
outflow model suppresses star formation in halos with circular velocities 
below 75$\kms$ more strongly than the previous vzw model~\citep{dav13}, 
delaying the 
early stages of reionization.  In short, the differences between our previous 
and current simulations generally delay the current model's reionization 
and suppress $\tes$.  This reinforces the well-known problem that 
reionization via galaxy formation, while possible within the concordance 
cosmology, is challenging.  

Despite the somewhat delayed reionization history, we believe that our 
model gives a plausible representation model for the UVB and metal absorbers
during the latter stages of the reionization epoch for two reasons.  First, 
the volume-averaged neutral hydrogen fraction drops below 1\% at $z=6.5$.
This is reasonably consistent with the observation that it is very small 
($<10^{-3}$) by $z=6$~\citep{fan06,bec14}, although large-scale fluctuations in 
the UVB may remain until $z=5$~\citep{sch13,bec14,mal14}.  Furthermore, we 
will show in Figure~\ref{fig:tau_v_z} that the simulated Lyman-$\alpha$ 
forest optical depth is either at the lower or upper end of the observed 
range, depending on our treatment of the UVB.  Hence although 
the model may not account for all ionizing sources at $z\geq10$, we expect 
that it does at later epochs.  Second, our mass resolution is higher than 
the w8n256 simulation discussed by~\citet{opp06}, which was already shown 
to yield converged predictions for absorbers down to columns of 
$10^{12}$cm$^{-2}$.  This indicates our simulation fully accounts for
the galaxy population that generates low-column absorbers.

\section{The Simulated Radiation Field}\label{sec:J}
In this section, we discuss our simulated radiation field.  Our model 
predicts that $\civ$ absorbers are photoionization-dominated because 
gravitational and outflow-driven shocks do not heat enough enriched 
gas to the temperatures at which $\ciii$ can be collisionally 
ionized (ODF09).  This is consistent with the observational result 
that, at $z\sim3$, $\civ$ absorbers in high-resolution spectra have 
narrow velocity widths implying an origin in 
photoionization~\citep{pro97,tes11}.  

At $z>3$,~\citet{cen11} predict a transition from photoionization to 
collisional ionization of $\civ$-absorbing gas.  Unfortunately, 
observations of $\civ$ absorbers at these redshifts cannot yet test 
whether they consist of multiple narrow ($b$-parameters $<25\kms$) 
components, which is necessary to rule out the collisional ionization 
hypothesis.  At $z\sim6$,~\citet{bec11} use high-resolution spectra 
of seven low-ionization absorbers to find velocity widths of 
10--100$\kms$~\citep{bec11}.  Four of these have velocity widths 
$\leq25\kms$, indicating negligible collisional ionization of CIII.  
The other three have widths of up to $200\kms$.  Such broad widths
indicate sufficiently energetic gas to collisionally ionize metals
completely.  Remarkably, however, even these systems are undetected 
in $\civ$ and $\siiv$ (with the possible exception of a weak $\civ$ 
system at $z=5.79$ in the spectrum of SDSS J0818+1722), hence the 
presence of low-ionization systems may indicate turbulent 
broadening of cold gas.  One of their quasars 
(SDSS J1030+0524) does contain coincident $\cii$ and $\civ$ 
absorption at $z=5.7425$, but this is only seen in the X-shooter 
spectrum of~\citet{dod13}, where the data quality do not permit 
a robust constraint on its $b$-parameter.  More high-resolution
data will be required to test the hypothesis that $\civ$
is collisionally-ionized at $z>3$. 

In short, an origin in photoionization remains consistent with 
available constraints at $z\geq3$.  For the present, we therefore
persist in the simple approach of asking how well our current 
simulation performs and speculate for the purposes of future work 
as to how important the subgrid scales are.

\subsection{Amplitude}\label{ssec:Jamp}
\begin{figure}
\centerline{
\setlength{\epsfxsize}{0.5\textwidth}
\centerline{\epsfbox{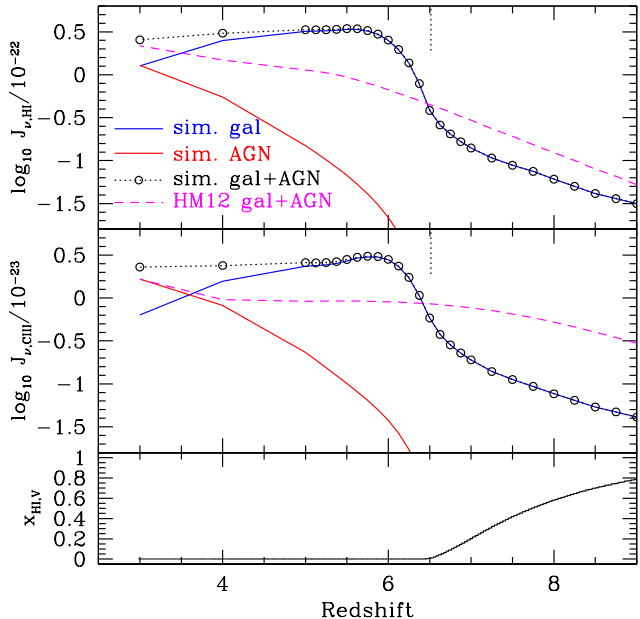}}
}
\caption{The amplitude of the ionizing background in ergs s$^{-1}$ 
cm$^{-2}$ Hz$^{-1}$ Sr$^{-1}$ at the ionizing edges
of $\hi$ \emph{(top)} and $\ciii$ \emph{(middle)}.  Red and blue solid 
curves show the contributions from galaxies and AGN, while the black 
dotted curve gives their total.  The dashed curve is from HM12, smoothed 
to the same spectral resolution as our simulation.  The bottom panel
shows the volume-averaged neutral hydrogen fraction $\xhiv$, while the 
vertical sements in the top and middle panels indicate where $\xhiv$ 
drops below 1\%.  The simulated radiation field's amplitude is 
2--3$\times$ larger than in the HM12 model throughout the observed 
epoch and is dominated by galaxies rather than AGN.
}
\label{fig:Jamp}
\end{figure}

In Figure~\ref{fig:Jamp}, we show how the simulated UVB's amplitude at 
the $\hi$ and $\ciii$ ionization edges evolves with time.  For comparison,
we also show the volume-averaged neutral hydrogen fraction in the bottom
panel.  Prior to the completion of hydrogen reionization, the 
volume-averaged UVB grows smoothly as the mean free path to ionizing photons 
increases.  Its amplitude then jumps when reionization completes ($z=6.5$).
This is a well-known numerical artefact of computing reionization in 
small volumes~\citep{bar04,ili14}, and owes to an overly-rapid increase 
in the mean free path of ionizing photons.

Following reionization, the contribution from galaxies to the UVB declines
slowly while the contribution from AGN increases more rapidly.  The combination
of the two yields a nearly invariant UVB amplitude, in qualitative agreement 
with observations~\citep{bec13}.  To our knowledge, this represents the first 
attempt to model both the pre- and post-reionization UVB in a three-dimensional
framework.  The fact that, despite its limitations, our model already reproduces 
the UVB's observed non-evolution is an encouraging victory.

For reference, we also show the UVB amplitude predicted by 
the HM12 model.  For a fair comparison with our simulation's low 
spectral resolution, we average the HM12 model over the energy 
range 1.0--1.5625 Ryd in the top panel and 3.25--3.8125 Ryd in the middle
panel.  The simulated background lies below the HM12 model prior 
to reionization, but this is not a serious problem as the latter is a 
volume-averaged calculation that becomes inaccurate in epochs where
the UVB becomes highly inhomogeneous.  Following reionization, the
simulated UVB is 2--4$\times$ stronger than the HM12 model, 
with the gap declining to $z=3$.  This discrepancy is not large compared
to the observational uncertainties, which characteristically span a 
factor of 2--3~\citep{bec13}.  Moreover, it is difficult to avoid in
numerical simulations because the UVB amplitude $J$ varies with the 
ionizing emissivity $\epsilon$ as 
$J\propto\epsilon^3$--$\epsilon^4$~\citep{mcq11}.  As long as galaxies
dominate the UVB, $\epsilon\propto\fesc$, so a factor of three in $J$
translates to a factor of $<1.5$ in $\fescfive$.  This nonlinear 
dependence, combined with the high computational expense of radiation
transport simulations, means that ``hitting the target" is very 
challenging.  For the present, we simply adjust our simulated UVB by 
a factor $J_{\rm{HM12}}/\langle J_{\rm{sim}}\rangle$ (where the amplitudes
are evaluated at the $\hi$ ionization edge) wherever it overproduces 
the HM12 model.  This calibration preserves the simulated 
UVB's shape and spatial inhomogeneity below $z=6.5$ and leaves it 
entirely unchanged at earlier times.  In what follows, we will refer 
to this UVB as the ``renormalized" one as compared to the directly-computed,
``simulated" one.

\begin{figure}
\centerline{
\setlength{\epsfxsize}{0.5\textwidth}
\centerline{\epsfbox{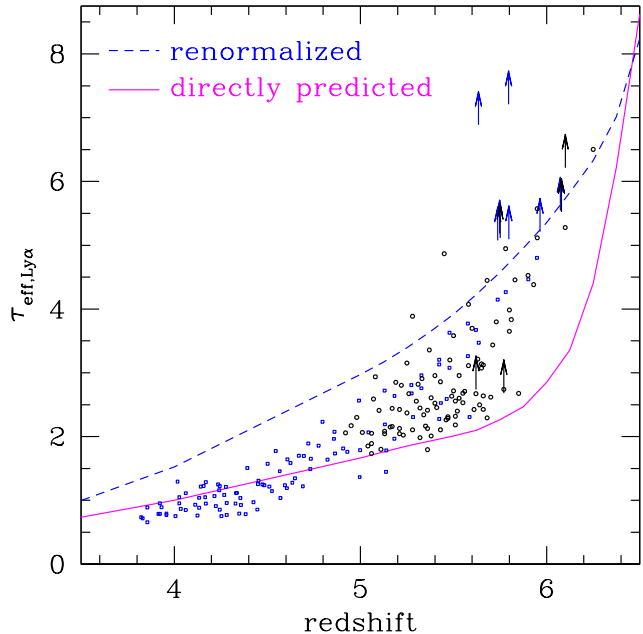}}
}
\caption{The volume-averaged optical depth to aborption by the Lyman-$\alpha$
forest as a function of redshift directly predicted by the simulation (solid 
magenta) and under the assumption of ionization equilibrium with the 
renormalized UVB (dashed blue).  The two curves broadly bracket the 
observed range at $z>5$ while the simulated curve is in excellent 
agreement with observations at later times.  Black circles and blue squares
are from~\citet{fan06} and~\citet{bec14}, respectively, while arrows indicate
regions where no transmission was detected at $2\sigma$.
}
\label{fig:tau_v_z}
\end{figure}

In order to examine to what extent the simulated and renormalized UVBs
agree with IGM observations, we have computed the volume-averaged optical 
depth to absorption by the Lyman-$\alpha$ forest $\tau_{\mathrm{Ly}\alpha}$ 
in both cases.  The solid magenta curve in Figure~\ref{fig:tau_v_z} shows 
the simulated optical depth.  It is in remarkably good agreement with 
observations for $z<5$~\citep{fan06,bec14}.  From $z=5$--5.8, it traces 
the lower end of the observed range, indicating that it is slightly too 
strong.  Above $z=5.8$, it is much stronger than implied by observations.

These discrepancies can be traced to two factors.  First, the tendency 
for the simulated $\tau_{\mathrm{Ly}\alpha}$ to evolve too quickly is
a well-known artefact of small volumes~\citep{bar04,ili14} that will
improve through increased dynamic range.  Second, the
Lyman-$\alpha$ forest is increasingly sensitive to rare voids at high
redshifts~\citep{bol09}; these are systematically absent from small
volumes.   Note that, by contrast, metal absorbers are likely weighted 
toward overdense regions where galaxies form.  In other words, our 
simulation probably yields a better model for metal absorbers than 
for the Lyman-$\alpha$ forest.

In order to produce the blue dashed curve corresponding to the 
renormalized UVB, we recomputed each gas particle's ionization state 
under the assumption of ionization equilibrium given the local 
temperature and density (this introduces 
$\sim1\%$ shifts in $\tau_{\mathrm{Ly}\alpha}$ with respect to a full 
non-equilibrium calculation).  The renormalized UVB traces the upper 
envelope of the observed range from $z=5-6$.  This is consistent with 
the view that our simulation lacks large-scale voids where the gas would
be on average more ionized~\citep{bol09}.  Expanding our dynamic range
would suppress $\tau_{\mathrm{Ly}\alpha}$ into improved agreement with 
observations.  

While Figure~\ref{fig:tau_v_z} does not obviously favour one UVB model
over the other, we will find it convenient to assume the
renormalized UVB in order to facilite comparisons with the HM12 model.  
At the same time, in Section~\ref{sec:obs} we will leverage the fact 
that the simulated and renormalized UVBs bracket observations to 
explore the impact of UVB fluctuations on length scales that our 
simulation does not capture.

\subsection{Spectral Slope}\label{ssec:Jslope}
\begin{figure}
\centerline{
\setlength{\epsfxsize}{0.5\textwidth}
\centerline{\epsfbox{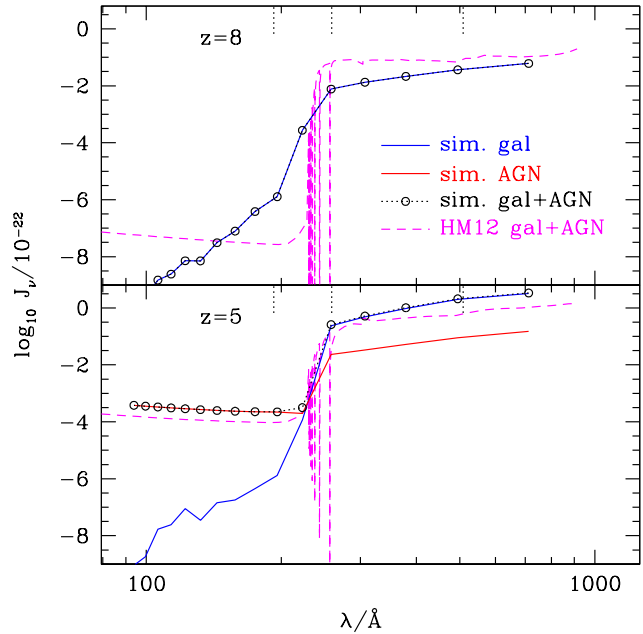}}
}
\caption{The simulated UVB from 1--10 Ryd at two representative
redshifts along with the HM12 model.  Dotted vertical ticks
indicate the ionization potentials of $\cii$, $\ciii$, and $\civ$.
Note that our model does not account for the sawtooth opacity of
singly-ionized helium owing to its low spectral resolution.
}
\label{fig:Jslope}
\end{figure}

We show in Figure~\ref{fig:Jslope} how galaxies and AGN contribute to the
total UVB at two representative redshifts and compare with the HM12 model.  
Note that these figures do not include the calibration discussed in 
Section~\ref{ssec:Jamp}.  This comparison reveals two differences in 
addition to the small amplitude offset discussed in 
Section~\ref{ssec:Jamp}.  First, we note that, redwards of the HeII 
ionization edge (228 \AA), the simulated background is redder at $z=8$ 
than in the HM12 model.  This discrepancy cannot owe to the metallicity
of the star-forming gas because the simulation's star formation-rate 
weighted mean metallicity is roughly 0.5--0.6 times as large as assumed
by HM12 (their equation 52) for $z=6$--10, hence we would have
expected a bluer continuum.  The difference probably owes to the dramatically
different ways that the two models treat the opacity from a partially-neutral 
universe.  In any case, it largely disappears by $z=5$, when it could
be probed observationally, hence we will not consider it further.

Second, the simulation yields a very different mean UVB bluewards 
of 228 \AA~for $z>6$ partly because we do not ``turn on" AGN until $z=8$, 
and partly because we do not assume
that $\fesc=0$ for photons with energies above 4 Ryd.  This allows galaxies
to ``jump-start" $\heii$ reionization and may eventually be testable via
observations of high-ionization metal absorbers at high redshifts, perhaps
along sightlines to gamma-ray bursts.  For the present, however, we have 
found that including or excluding the galaxy flux with energies greater than 
4 Ryd does not affect the predicted abundance of $\cii$ and $\civ$ absorbers, 
hence the two models are equally acceptable.

\subsection{Scatter}\label{ssec:Jscat}
\begin{figure}
\centerline{
\setlength{\epsfxsize}{0.5\textwidth}
\centerline{\epsfbox{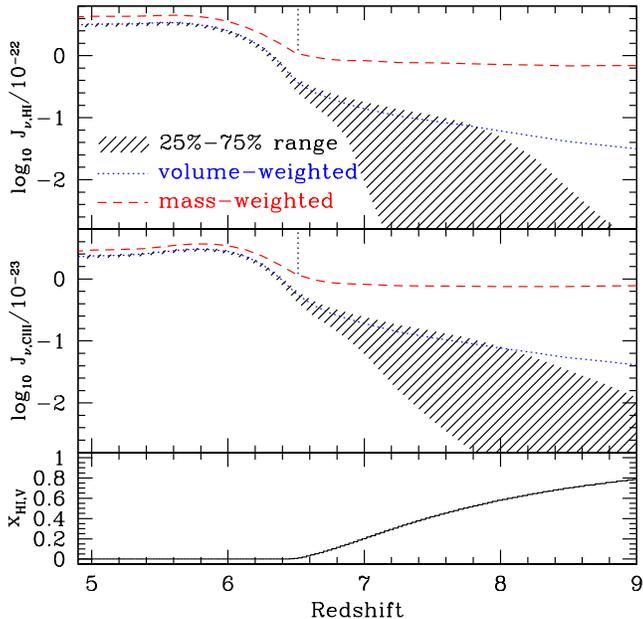}}
}
\caption{The scatter in the simulated radiation field at the
HI \emph{(top)} and $\ciii$ \emph{(middle)} ionization edges.  For
reference, we include the volume-averaged neutral fraction in the
bottom panel and use vertical segments to indicate the reionization
redshift as in Figure~\ref{fig:Jamp}.  The UVB amplitude varies by 
orders of magnitude
prior to the completion of reionization.  When considering only
the galaxy contribution, the mass-weighted UVB is always stronger 
than the volume-weighted UVB, which suggests that the 
volume-averaged UVB underestimates the field that metal absorbers 
experience by $>30\%$ for $z>4$.
}
\label{fig:Jscat}
\end{figure}
As observations push into the reionization epoch, the UVB becomes
increasingly inhomogeneous.  A key strength of our model is its
ability to capture this inhomogeneity down to length scales of 30--40
proper kiloparsecs, which is roughly half the distance out
to which metals travel ($\sim100$ proper kpc;~\citealt{opp08}).  Given
that galaxies cluster on much larger scales, our 
simulation is therefore beginning to resolve the impact of the local 
field on metal absorbers.  In order to assess whether this could play
a role in our predictions, we show in Figure~\ref{fig:Jscat} the 
interquartile range as well as the volume-weighted and mass-weighted 
UVB ($J_V$ and $J_M$, respectively) at the $\hi$ and $\ciii$ ionization edges as a 
function of redshift.  To compute $J_M$, we weight the UVB in each
radiation transport voxel by the amount of collapsed mass that it
contains; this presumably provides a fair representation of the field
near virialized regions.  Note that Figure~\ref{fig:Jscat} does not 
include the calibration discussed in Section~\ref{ssec:Jamp} as doing 
so would obviously leave results regarding spatial fluctuations 
unchanged.

Not surprisingly, the UVB fluctuates enormously prior to the epoch of 
overlap (we define this is the redshift where the volume-averaged
neutral hydrogen fraction drops below 1\%; this occurs at $z=6.5$ in
our simulation).  What is perhaps surprising is how quickly it becomes
inhomogeneous: already if the neutral fraction is only 10\% (which it
may be at $z=7$;~\citealt{bol11,bh13}), the interquartile range spans a factor of 2.5 (2) 
at the $\hi$ ($\ciii$) ionization edges.  The obvious implication is that 
it is inappropriate to model metal absorbers with a homogeneous 
UVB at redshifts where the neutral fraction is greater than 
$\sim10\%$; the local field will dominate the photoionization 
rate.

Following reionization, fluctuations are at the $\sim10\%$ level.
Interestingly, $J_M > J_V$ at all times, and the difference is larger
than the interquartile range.  Even at $z=5$, when the mean free path to
ionizing photons
is much larger than our simulation volume~\citep{wor14}, $J_M/J_V$
is 1.3 (1.2) at the $\hi$ ($\ciii$) ionization edges.  This means that,
as long as the UVB is not dominated by AGN, it is not possible for
models to predict the abundance of metal absorbers to better than
this accuracy unless they take local-field effects into account.  
Importantly, this is true even for optically-thin ions that are 
seen out to large impact parameters (such as $\civ$).  This is in
contrast to the Lyman-$\alpha$ forest, which is relatively insensitive
to the large spatial fluctuations that exist at $z>5$ because it
is dominated by gas that lives in overall less biased 
regions~\citep{mes09}.

\section{Comparison With a Volume-Averaged UVB}\label{sec:compHM12}
The most important differences between our current simulation and
previous work are (1) The introduction of a spatially-resolved UVB;
and (2)
A subgrid treatment for self-shielding that explicitly attenuates the 
UVB in dense regions.  In this section, we compare where absorbers lie
in temperature-density space within our model versus expectations from 
the HM12 background in order to indicate what physical conditions
$\cii$ and $\civ$ absorbers trace in each case.  We then show how the 
overall ionization state of carbon evolves in order to motivate a 
discussion of the evolving abundance of $\cii$ and $\civ$ absorbers.

\begin{figure}
\centerline{
\setlength{\epsfxsize}{0.5\textwidth}
\centerline{\epsfbox{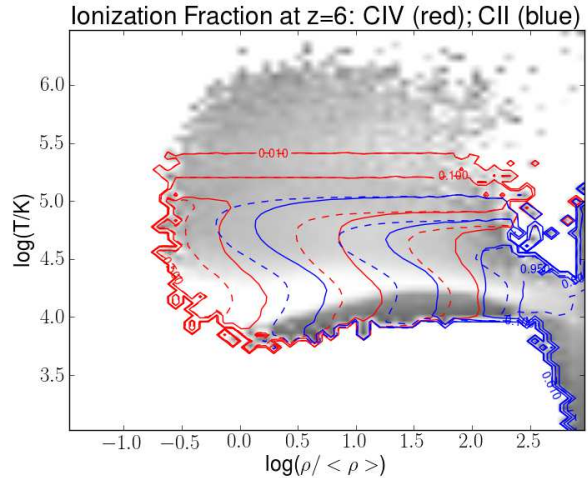}}
}
\caption{Solid red and blue contours indicate the mass fraction of carbon
that is in $\civ$ and $\cii$ as a function of density and temperature
at $z=6$ using our simulated UVB with the normalization discussed
in Section~\ref{sec:J}.  Dashed contours are the same, but adopting the
spatially-uniform HM12 UVB.  Gray shading is proportional to
the logarithm of the mean carbon mass fraction.  Broadly, the simulated 
UVB yields significantly more ionized gas even though the two 
backgrounds are calibrated to match at the hydrogen ionization edge.
}
\label{fig:phaseSpace}
\end{figure}

We show in Figure~\ref{fig:phaseSpace} how the mass fractions of carbon
in $\cii$ and $\civ$ (blue$=\fcii$ and red$=\fciv$, respectively) vary 
with temperature and density at $z=6$ in the presence of our 
(spatially-inhomogeneous) renormalized UVB (solid) as compared to the 
spatially-uniform HM12 UVB (dashed).  Broadly,
the behaviour is similar to the results presented in ODF09: $\civ$ 
prefers gas whose density is at or below the mean and has temperatures 
below $10^5$K, indicating an origin in photoionization.  Meanwhile, 
$\cii$ prefers gas with densities more than $30\times$ the cosmic 
mean and temperatures below 50,000 K.

Comparing the dashed and solid curves, we see that our simulated 
UVB pushes the contours to higher density with respect to 
the HM12 model.  In other words, gas is more highly ionized
at all densities less than $\sim100$.  The generally higher 
ionization state reflects the impact of the local radiation field.
At higher densities, self-shielding boosts the model's $\cii$ 
fraction with respect to expectations from the HM12 UVB.  

These differences have implications for the relationship between galaxies 
and absorbers.  ODF09 find that, during the reionization epoch, $\civ$ 
traces the locations of relatively massive galaxies because only 
they can expel metals out to the low densities where $\civ$ becomes the 
dominant ionization state.  Meanwhile, $\cii$ prefers higher densities 
and hence dominates in regions that are closer to the originating 
galaxies, and since low-mass galaxies dominate by number, they 
host most of the $\cii$ absorbers.  Given that the local UVB boosts 
the ionization parameter near halos, it increases the $\civ$ mass 
fraction and suppresses $\cii$ at all masses.

\begin{figure}
\centerline{
\setlength{\epsfxsize}{0.5\textwidth}
\centerline{\epsfbox{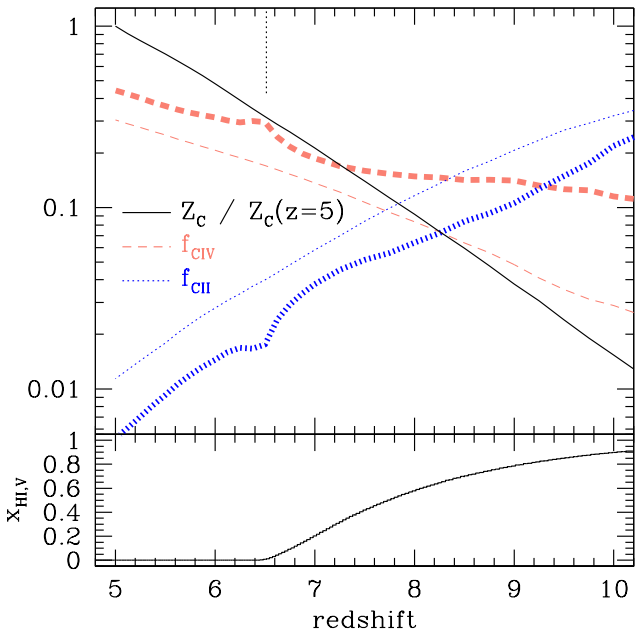}}
}
\caption{The evolution of the IGM's carbon metal mass fraction $Z_c$ 
(normalized to its value at $z=5$) and the volume-weighted mass 
fractions that are in $\civ$ and $\cii$.  The bottom panel shows
the simulated reionization history while the vertical segment in the
top panel indicates the simulated reionization redshift.  Heavy and 
light curves show results from our simulated UVB and the HM12 UVB, 
respectively.  
From $z=5$--10, both $Z_C$ and $\fciv$ increase while $\fcii$ 
decreases.  Meanwhile, the simulation predicts generally more ionized
gas than the HM12 UVB owing to the local field from galaxies.  
This figure clearly shows that observations need to transition to 
low-ionization ions as they approach the reionization epoch.
}
\label{fig:Cfracs}
\end{figure}

In Figure~\ref{fig:Cfracs}, we elaborate on this point and move a step 
closer to observational implications by showing 
how the volume-weighted mass fractions of carbon in $\cii$ and $\civ$
evolve along with the IGM carbon metal mass fraction $Z_C$, normalized to its
value at $z=5$.  Here, we define as ``IGM" any gas that is not star-forming.
Note that the normalizations of these curves are different  from those of 
the total ion abundances $\ocii$ and $\ociv$ (which we will explore in 
Figure~\ref{fig:Omega_predict}) because the latter are weighted to high 
cross sections and columns.  Nonetheless, they give qualitative insight 
into the expected observational trends.  For reference, we include the 
volume-averaged neutral hydrogen fraction in the bottom panel.  

Focusing on the heavy curves
first, our simulation predicts that $\fcii$ declines from 20--30\% at 
$z=10$ to $<1\%$ by $z=5$.  Over the same interval, $\fciv$ grows from 
10\% to 50\%.  In other words, $\fcii$ and $\fciv$ essentially
``swap places" during this epoch.  It follows that observations should
transition from high-ionization to low-ionization absorbers as they 
approach the reionization epoch, as that is where most of the carbon
is.  We will discuss how the abundance of $\cii$ and $\civ$
absorbers is predicted to evolve at $z>5$ in Section~\ref{sec:predict}.

The thin dashed and dotted curves show what is expected if we replace our
renormalized UVB with the HM12 model.  Despite the fact 
the two backgrounds match at the hydrogen ionization edge, the simulated 
UVB predicts significantly more $\civ$ and less $\cii$.  This reflects
the fact that the UVB experienced by metals is stronger than
the volume-averaged field.  It is interesting to note that self-shielding
does not cancel the impact of the the local field on $\cii$.  This
is because it boosts the $\cii$ mass fraction predominantly in highly
overdense regions (Figure~\ref{fig:phaseSpace}) whereas 
Figure~\ref{fig:Cfracs} is weighted toward lower densities.

The behavior in Figure~\ref{fig:Cfracs} is in qualitative agreement 
with the observation by~\citet{bec11} that their observed 
sample of low-ionization systems was skewed to the higher-redshift 
portion of their survey.  In the next section, we map 
Figure~\ref{fig:Cfracs} into observable spaces and ask how its 
predictions could be tested.

\section{Comparison with Observations}\label{sec:obs}
In this section, we study the predicted abundance of $\cii$ and $\civ$
absorbers.  We will show that our simulation reproduces the observed
abundance of $\civ$ and $\cii$ absorbers at $z=6$ in the regime where 
the simulated and observed dynamic ranges overlap, and at $z=3$ for
$\civ$ absorbers with $\lognciv>14$.

\subsection{Generating Simulated Spectra}\label{ssec:ana}

We begin by discussing how we extract simulated spectra from our model
and compute the abundance of absorbers.  At each redshift, we cast
a sightline that is oblique to the simulation boundaries and wraps around
it until subtending a Hubble velocity width of $5\times10^5\kms$.  
This corresponds to an absorption path length~\citep[e.g., Equation 2 of][]{bec11} 
of 20--100, depending on the
redshift.  The line of sight is divided into bins whose width $\Delta$ 
in configuration space corresponds to a separation of $2\kms$ in a pure 
Hubble flow; that is, $\Delta \times H(z) = 2 \kms$.
We smooth particles that overlap the line of sight to the radiation
field's grid in order to compute the local ionization rates of all ions
of O, C, and Si.  Next, we compute the equilibrium ionization state of
these atoms using the local temperature, density, metallicity, and UVB (where
we emply the renormalized UVB unless stated otherwise).
Our ionization solver accounts for 
photoionization, collisional ionization, direct and dielectronic recombination, 
and charge transfer recombination.  Photoionization cross-sections are taken 
from~\citet{ver96}; collisional photoionization rates are from~\citet{vor97}; 
direct and dielectronic recombinations are from~\citet{bad06} 
and~\citet{bad03}, respectively; and charge transfer recombination rates
are from~\citet{kin96}.  The local hydrogen ionization
state is fixed to the nonequilibrium value stored in the simulation snapshot,
and the electron abundance is taken as the value from the simulation owing
to ionization of hydrogen and helium plus the small contribution from ionized 
metals.  This approach is essentially equivalent to 
running {\sc CLOUDY}~\citep{fer98} on each gas particle because we use the same 
cross-sections and rates except that we neglect Auger ionization. 

We derive each transition's optical depth along the sightline as a function 
of velocity following a standard approach that takes thermal and bulk motions 
into account~\citep[for example,][]{the98}.  We then smooth the predicted
transmission with a Gaussian kernel whose full-width at half maximum is
$10\kms$ in the rest-frame and add random noise corresponding to a 
signal-to-noise of 50 per pixel.  Finally, we use {\sc autovp}~\citep{dav97} 
with the default 
parameters to identify absorption systems and compute column densities.

In order to make a fair comparison to observations, we follow standard
practice and merge all absorbers that lie within $50\kms$ of each other 
into ``systems" and add their column densities.  In what follows, we
will will use the terms ``systems", ``absorbers," and ``absorption
systems" interchangeably.

\subsection{Column Density Distribution}\label{ssec:cdd}

\begin{figure}
\centerline{
\setlength{\epsfxsize}{0.5\textwidth}
\centerline{\epsfbox{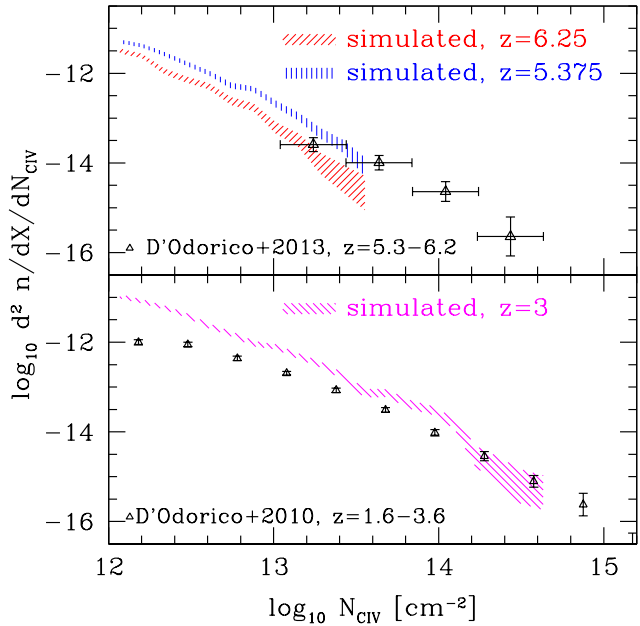}}
}
\caption{Simulated column density distribution versus observations at 
$z\approx5$--6 (top panel) and $z=1.6$--3.6 (bottom panel).  The simulated 
range reflects sqrt(N) uncertainty.  We have
multiplied the~\citet{dod13} measurements by 1.07 and the~\citet{dod10}
measurements by 1.197 to adjust to our cosmology. The predicted
abundance near the end of the reionization epoch shows good agreement 
with observations in the column density range $\lognciv=$13--13.5, with 
an abundant population of absorbers predicted at lower columns.  By 
$z=3$, the simulation overpredicts the observations,
particularly at the low-column end.
}
\label{fig:dNdX_CIV}
\end{figure}

In the top panel of Figure~\ref{fig:dNdX_CIV}, we  compare the simulated column 
density distribution (CDD) of $\civ$ absorption systems at $z\sim6$ versus the 
observations of~\citet{dod13} at $z=5.3$--6.2, where we have adjusted their 
measurements to match our assumed cosmology.  We find reasonable agreement 
in the range $\lognciv=$13--13.5.  This suggests that the assumed carbon 
yields and outflow scalings at the reionization epoch are broadly realistic.  
Unfortunately, the simulation underproduces systems with larger columns.
This reflects a well-known numerical artefact whereby small simulation
volumes whose mean matter density equals the cosmic value systematically
lack the massive systems that dominate $\civ$ absorbers at high redshift.  

Our high mass resolution allows us to predict the CDD down to lower columns 
than previous works~\citep[such as][]{cen11}; this is the tradeoff for our 
much smaller simulation volume.  We find that the predicted CDD remains 
quite steep down to at least $10^{12}$ cm$^{-2}$, as directly observed at 
lower redshifts (bottom panel).  Future observations probing to lower 
columns will therefore impose stronger constraints both on our newest 
outflow model and, more generally, on reionization models in which 
galaxies in photosensitive halos dominate the ionizing background.

In the bottom panel, we compare the predicted column density distribution 
at $z=3$ versus the~\citet{dod10} observations at $z=1.6$--3.6.  At
high columns ($\lognciv > 14$), the predicted CDD lies within 
$1\sigma$ of observations.  At lower columns, the model overpredicts 
observations by 2--3$\times$, which is much larger than the measurement
uncertainty.  Note that observational completeness is not
expected to be the main problem:~\citet{dod10} estimate a completeness 
of 60\% at $10^{12}$cm$^{-2}$ and $>90\%$ for $\lognciv > 12.6$.

If the discrepancy owes to the simulated UVB, then it must owe either 
to its slope or to spatial fluctuations rather than the amplitude,
which is reasonably well-constrained at $z=3$ (recall that we calibrate
it to match the HM12 model at the $\hi$ ionization edge).
Obviously, the tendency 
for the UVB to be stronger near galaxies contributes, but only at the 
30\% level (Section~\ref{ssec:Jscat}).  Moreover, we believe this is a 
real effect, so it cannot be the true problem.  An alternative explanation 
is that our model underestimates the UVB at the $\civ$ ionization edge 
owing to delayed HeII reionization (Section~\ref{ssec:Jslope}); in other 
words, the model underestimates the mass fraction that is photoionized
to $\cv$ and higher ionization states because the opacity bluewards 
of 4 Ryd is too strong.

\begin{figure}
\centerline{
\setlength{\epsfxsize}{0.5\textwidth}
\centerline{\epsfbox{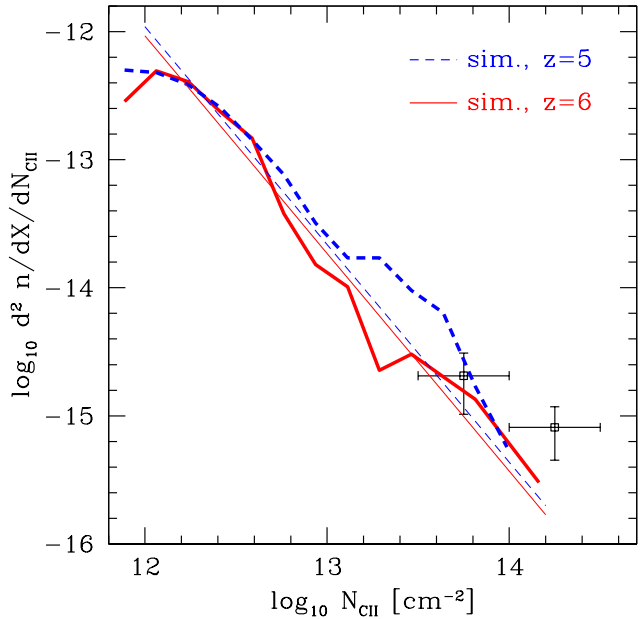}}
}
\caption{Simulated $\cii$ CDD at two different redshifts as indicated.
Data are derived from 9 observed $\cii$ systems at absorption redshifts
of 5.57--6.25 (see text), with vertical errors indicating sqrt(N).  Thin
curves indicate best-fit power laws assuming a constant slope of -1.7.
The simulation reproduces observations within the errors, indicating
that the simulated CGM and radiation fields are broadly realistic.
}
\label{fig:dNdX_CII}
\end{figure}

It is also possible that the model overproduces $\civ$ because it 
ejects too many metals into the IGM.  We disfavor this interpretation
because, as we will show in Figure~\ref{fig:dNdX_CII}, our model
reproduces the observed abundance of $\cii$ absorbers.  Suppressing 
outflows in order to reduce the number of $\civ$ absorbers would 
cause the model to underproduce $\cii$ severely.

Finally, we note that, if numerical limitations are any problem at all 
at $z=6$, then they are a much bigger one at $z=3$ because small
simulation volumes miss a progressively larger fraction of collapsed
structure at lower redshifts~\citep{bar04}.  At a glance, it is therefore 
quite surprising that our simulation extends to higher columns
at $z=3$ than at $z=6$.  The explanation may be found in the tendency 
for $\civ$ to trace higher overdensities at lower 
redshifts~\citep[ODF09;][]{dod13}.  The higher the characteristic
overdensity of an ion, the more rapidly galaxies can enrich their
surroundings to the point where that ion becomes observable.  
Consequently, limitations associated with our small simulation volume 
are less severe for $\civ$ at lower redshifts, particularly when 
considering absorbers in a fixed range in column density.

In Figure~\ref{fig:dNdX_CII}, we compare the predicted $\cii$ CDD to
observations.~~\citet{bec11} identified 7 $\cii$ systems in high-resolution
spectra of 9 quasars that have a mean emission redshift of 6; 6 of these
have column densities above $10^{13.5}$cm$^{-2}$, where they report
more than 50\% completeness.~~\citet{dod13} identified 5 $\cii$ systems
in lower-resolution spectra of 6 quasars whose mean emission redshift is 5.7.
Three of these quasars and two of the $\cii$ systems were also observed at 
higher resolution by~\citet{bec11}.  We combine these samples, adopting 
the~\citet{bec11} column densities for the two 
absorbers in the spectrum of SDSS J0818+1722.  This yields a total of 9 $\cii$ 
systems with column densities above $10^{13.5}$cm$^{-2}$.  We assume that each
sightline surveys an absorption path length of 2.37, for a total surveyed
path length of 28.44.  Our simulation reproduces the observed $\cii$ CDD within
the errors.

In summary, the level of agreement between the predicted and observed $\cii$ and
$\civ$ CDDs in Figures~\ref{fig:dNdX_CIV}--\ref{fig:dNdX_CII} suggests that the
metallicity, temperature, and radiation field of the reionization-epoch CGM
in our model are realistic.  The fact that we have not tuned any parameters
in order to match these observations supports the view that current observations
of metal absorbers do not require additional physical inputs such as metals
or light from Population III stars~\citep[see also][]{kul13} or a strong 
contribution from X-ray binaries to the UVB~\citep{jeo14}.

\section{Predictions}\label{sec:predict}
\subsection{Integrated Abundances}\label{ssec:abun}
In this section, we predict the integrated abundance of $\cii$ and $\civ$ 
absorbers 
as a function of redshift.  In order to correct for the limitations
that we identified in Figures~\ref{fig:dNdX_CIV} and~\ref{fig:dNdX_CII},
we extrapolate from our simulation as follows:  First, we assume
that the simulated $\civ$ and $\cii$ CDDs are complete over the
column density range $10^{12}$--$10^{13}$cm$^{-2}$.  Next, we assume
that they follow power-laws with slopes of -1.7 (where the slope for 
$\cii$ is inferred from Figure~\ref{fig:dNdX_CII} and the slope for 
$\civ$ is supported by~\citealt{dod13}), and use the predicted 
abundances of systems with columns in the resolved range to 
normalize.  Finally, we compute the predicted number of systems with
column densities in the range $10^{13}$--$10^{15}$cm$^{-2}$ per
absorption path length as a function of redshift.  For reference,
we also include the expected number of systems in the expanded
range $10^{12}$--$10^{15}$cm$^{-2}$ as the lower-column systems will
be observable using the next generation of thirty-meter telescopes.

\begin{figure}
\centerline{
\setlength{\epsfxsize}{0.5\textwidth}
\centerline{\epsfbox{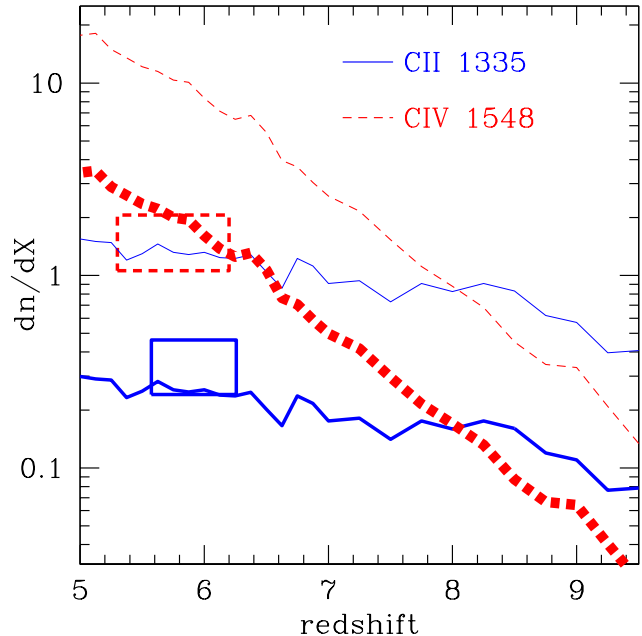}}
}
\caption{The predicted number of $\cii$ (solid blue) and $\civ$ 
(dashed red) systems per unit absorption path length as a function of 
redshift.  Heavy curves indicate the expected number of systems in the 
column density range $10^{13}$--$10^{15}$cm$^{-2}$, whereas thin curves
correspond to a next-generation survey encompassing 
$10^{12}$--$10^{15}$cm$^{-2}$.  All curves include an extrapolation
as described in the text.  The predicted number of $\cii$ systems evolves
slowly while the predicted $\civ$ abundance decreases strongly with 
increasing redshift.  The solid blue and dashed red boxes indicate
$\cii$ and $\civ$ observations, respectively.
}
\label{fig:dndX_predict}
\end{figure}

In Figure~\ref{fig:dndX_predict}, we show how the total number of systems 
per path length is predicted to evolve with redshift.  Perhaps the 
most important takeaway from this figure is what is \emph{not} seen, 
namely a sharp feature in the abundance of $\cii$ and $\civ$
systems at the reionization redshift ($z=6.5$).\footnote{In detail, 
there is a blip at $z=6.5$ that is particularly noticeable
in the $\cii$ curves, but this is a numerical artefact of the fact that
the simulated UVB grows rapidly in small simulation volumes.}
This reflects the fact that carbon
is ionized by local sources long before reionization is complete.  Going 
to lower column densities does not help because lower column densities
are populated by systems with lower masses, which are quite abundant.
Hence our model indicates that metal absorbers are sensitive both to
ongoing enrichment and to reionization.  We will return to this point
in Figure~\ref{fig:whatEvolves}.

The abundances of $\cii$ and $\civ$ systems evolve in very different ways.  
In particular, the abundance of $\cii$ systems evolves very slowly owing
to a tradeoff between the competing effects of enrichment and 
ionization~\citep[as was also noted in ODF09 and][]{fin13}.
Meanwhile, the abundance of $\civ$ systems increases strongly with
decreasing redshift because the strengthening ionizing background, 
increasing metallicity, and decreasing overall gas densities all push
an increasing fraction of the IGM into the range favoring $\civ$.

For a constant minimum column
density, the abundance of $\civ$ systems eventually drops below the
$\cii$ abundance.  This is qualitatively consistent with the
observation that $\civ$ is not observed past $z=6$~\citep{bec11}, 
although Figure~\ref{fig:dndX_predict} predicts that $\civ$ 
absorbers vanish much more slowly.

The solid blue box indicates the observed $\cii$ abundance.
We compute this by taking the 10 $\cii$ systems from 12 sightlines 
observed by~\citet{bec11} and~\citet{dod13}, each of which is assumed 
to probe a path length of 2.37.  The vertical error is sqrt(N).  
This estimate is in marginal agreement with the prediction.  
Correcting the data for the known incompleteness over this 
column density range ($>50\%$ for 
$N_\cii<10^{13.5}$cm$^{-2}$;~\citealt{bec11}) would boost the
observed abundance and degrade the level of agreement.  
On the other hand, increasing our simulation volume would likewise 
boost the predicted abundance by sampling the mass function more 
completely~\citep{bar04}.  Given these uncertainties, we
conclude that the predicted $\cii$ abundance is consistent with
current observations.

The dashed red box indicates the observed abundance of $\civ$ absorbers
from $z=5.3$--6.2.  We compute 
the observed value by using a least-squares approach to fit power-laws 
to the measurements of~\citet{dod13} and extrapolating to encompass the 
range $10^{13}$--$10^{15}$cm$^{-2}$.  Although current observations do 
not constrain the power-law parameters tightly~\citep{dod13}, this 
extrapolation is fairly robust because it only strays slightly outside the 
observed range.  The resulting best-fit value and 67\% confidence interval 
is $1.3^{+0.8}_{-0.2}$ for a fiducial slope of -1.7.  This is in good
agreement with the simulation, particularly at $z\sim6$.
\begin{figure}
\centerline{
\setlength{\epsfxsize}{0.5\textwidth}
\centerline{\epsfbox{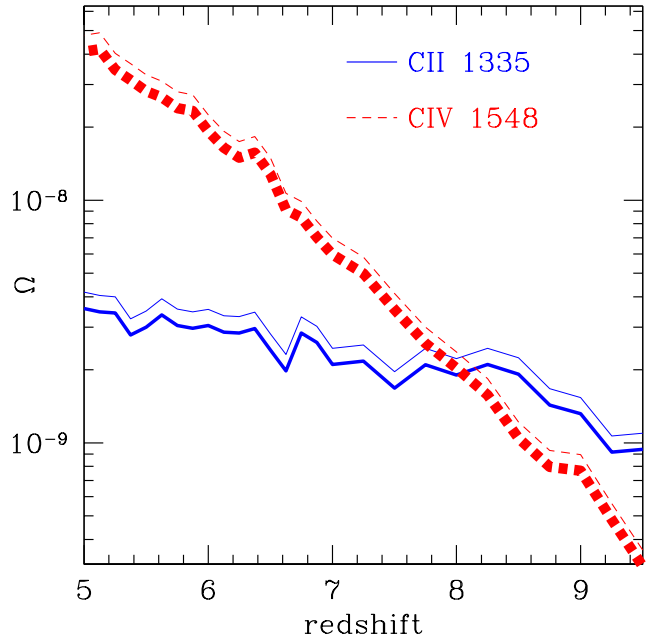}}
}
\caption{The predicted mass density in $\cii$ (solid blue) and $\civ$ 
(dashed red) systems as a function of redshift.  Heavy curves sum over
systems in the column density range $10^{13}$--$10^{15}$cm$^{-2}$, 
whereas thin curves correspond to a next-generation survey encompassing 
$10^{12}$--$10^{15}$cm$^{-2}$.
}
\label{fig:Omega_predict}
\end{figure}

We next show in Figure~\ref{fig:Omega_predict} how the predicted mass density in
each ion relative to the critical mass density evolves with redshift using
the same column density ranges as in Figure~\ref{fig:dndX_predict}.  
Broadly, the same trends are seen as before: $\ociv$ evolves rapidly whereas 
$\ocii$ evolves slowly, with the two crossing at $z=8$.  This may be slightly 
too early: As noted in Section~\ref{sec:intro}, observations indicate that
$\ocii > \ociv$ at $z=6$.  We attribute this discrepancy to uncertainty
in the CDD slopes, which are poorly-constrained at these redshifts.
Our assumption that both CDDs have power-law slopes of -1.7 affects 
Figure~\ref{fig:Omega_predict} more strongly than 
Figure~\ref{fig:dndX_predict} because it weights
the $\Omega$s toward the strongest absorbers, where the model is most
incomplete.  If, for example, the $\cii$ slope were flatter or the 
$\civ$ slope steeper, then they would cross at a later redshift.

In summary, our model reproduces the observed number of $\civ$ and $\cii$ 
absorbers per path length at $z\sim6$ within the uncertainties, and
assuming that both CDDs have slopes of -1.7 leads to the prediction that
$\ociv$ and $\ocii$ cross at $z\approx8$.  The former prediction is constrained
and reasonably robust, while the latter will need to be revisited using
calculations that treat a larger dynamic range in order to model the
CDDs' power-law slopes more precisely.

\subsection{Scatter}\label{ssec:abunScat}
Up until now we have considered the total number of absorption
systems, but we have not investigated the scatter in the number per
line of sight.  This could be significant for $\cii$ systems 
because there is roughly 1 per line of sight at $z=6$.  For 
example,~\citet{bec11} identify 7 systems along 9 sightlines.  Four 
of their systems lie along a single sightline while 6 sightlines have
no low-ionization absorption at all (although one of these 6, 
SDSS J1030+0524, does yield up a $\cii$ absorber in the X-shooter
spectrum  of~\citealt{dod13}).  Combining
the~\citet{bec11} and~\citet{dod13} samples, all of the observed $\cii$ 
systems are accounted for by only one third of their sightlines.  Does
this indicate that the Universe is partially neutral at $z=6$ or could 
it reflect ordinary galaxy clustering? These considerations motivate an 
inquiry into the scatter in the number of systems per observed line of 
sight or per survey.  Our simulated sightlines subtend absorption 
path-lenghths of 
20--100, while a single observed line of sight at $z\sim6$ can only probe 
$\cii$ systems over a path length of 2--3, hence we may use our simulation
to study the scatter in the predicted number of absorbers.  In practice,
this will underestimate the actual scatter owing to our small simulation
volume, but it is nonetheless a useful first step.

We begin by breaking up our simulated lines of sight into individual 
segments whose absorption path length corresponds to the redshift 
distance over which $\cii$ is observable in a quasar at that redshift.  
This is taken to be the distance between Lyman-$\alpha$ in the quasar 
rest-frame and 5000 $\kms$ bluewards of the $\cii$ 1334 transition 
(mimicking observational efforts to avoid the quasar's proximity zone).  
Note that, throughout this section, we use the simulated number of 
absorbers ``out of the box" rather than correcting for missing strong 
absorbers as in Figures~\ref{fig:dndX_predict}--\ref{fig:Omega_predict}.  
This will lead us to underestimate the predicted number of 
systems with columns greater than $10^{14}$cm$^{-2}$, but not have
much effect at lower columns.

\begin{figure}
\centerline{
\setlength{\epsfxsize}{0.5\textwidth}
\centerline{\epsfbox{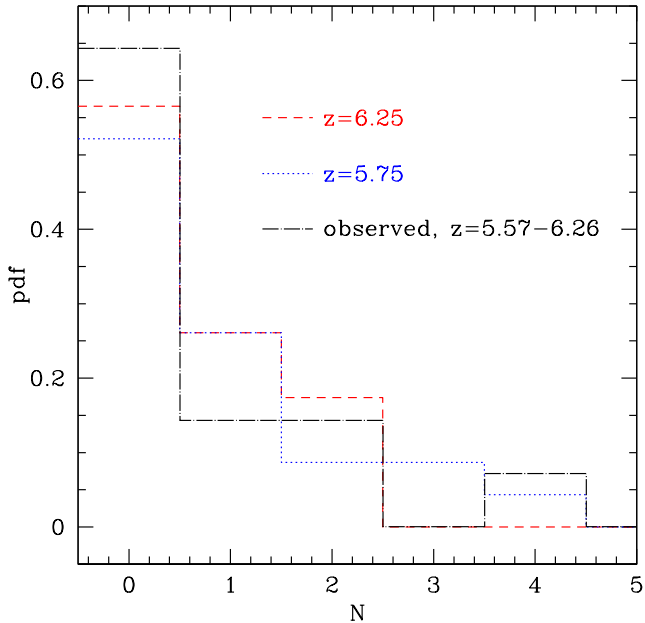}}
}
\caption{The distribution in the predicted number of $\cii$ 
systems with column density greater than $10^{13}$cm$^{-2}$
per sightline at two redshifts and in observations.  All histograms
are normalized to unit area.  Both the simulation and the observations 
find many sightlines with no systems at all and a small number of 
sightlines with one or more.
}
\label{fig:LOSScat}
\end{figure}

In Figure~\ref{fig:LOSScat}, we show the distribution in the predicted
number of systems per sightline at two representative redshifts and in
the observations.  For reference, the observations consist of 12 
independent sightlines while the simulation yields 23 at each 
redshift.  Broadly, the simulation reproduces both the tendency for a large
number of sightlines to contain no systems at all and an occasional 
sightline to contain more than one.  At the abundant end, three of 23 
simulated sightlines drawn from the $z=5.75$ snapshot contain
$\geq3$ systems (and none at $z=6.25$).  By comparison, one of 12 
observed sightlines contains 4 systems at $z=$6.13--6.25.  At the same 
time, more than half of sightlines have no absorbers both in the models 
and the data.  We conclude that, subject to our numerical limitations 
and the small current sample sizes, our simulation reproduces the 
observed clustering behavior of $\cii$ systems.  Given that our 
simulation volume completes reionization at $z=6.5$, this implies 
that $\cii$ observations are consistent with a completely reionized 
Universe.

\begin{figure}
\centerline{
\setlength{\epsfxsize}{0.5\textwidth}
\centerline{\epsfbox{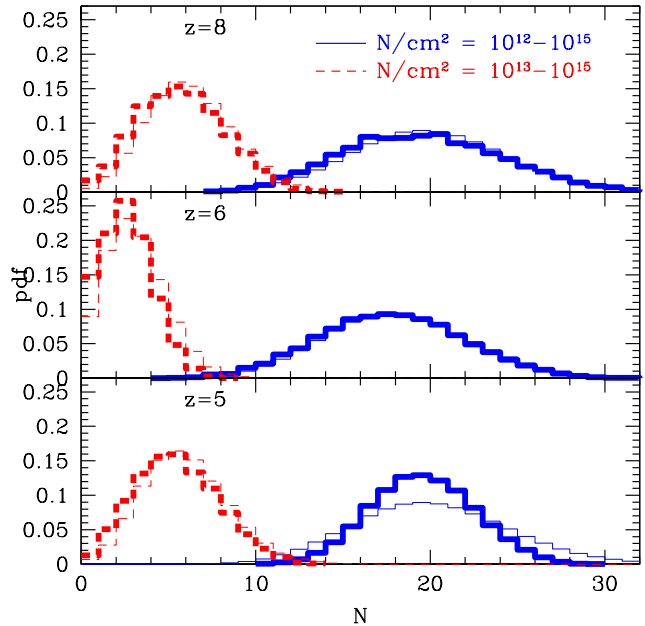}}
}
\caption{The distribution in the predicted number of $\cii$ systems 
observed in hypothetical surveys covering 6 quasars (or GRBs) at three 
different redshifts and two different column density ranges.  Heavy 
and thin curves indicate the raw simulated number of systems and 
Poisson distributions with the same means.  The predicted number of 
systems is skewed to low numbers with respect to a matching Poisson 
distribution.
}
\label{fig:compPois}
\end{figure}

Having studied the scatter in the number of absorbers per sightline, we
assemble 10,000 mock ``surveys" by drawing 6 simulated sightlines at random 
and compiling the number of absorbers in two ranges of column density.
In Figure~\ref{fig:compPois}, we use heavy curves to show the distribution 
in the total number of systems per survey at three redshifts and in two
column density ranges.  The mean number of systems per mock survey 
increases from $z=5$ to $z=8$ because the increasing path length per 
sightline more than cancels the declining intrinsic number of systems per 
path length (Figure~\ref{fig:dndX_predict}).  In fact, these effects
give rise to a local minimum in the predicted number of systems per path 
length which appears to occur somewhere between $z=8$ and $z=5$.  The effect
is not seen in Figure~\ref{fig:dndX_predict} because the latter extrapolates 
from the predicted number of systems with columns of 
$10^{12}$--$10^{13}$cm$^{2}$; it is therefore apparent only in stronger
systems.  Future work will be necessary in order to determine to 
what extent it reflects cosmic variance limitations.

For each combination of redshift and column density, we additionally use
a thin curve to show the Poisson distribution corresponding to the same 
mean number of systems.  In most cases, there are slightly more surveys with 
low numbers of systems than predicted by the Poisson distribution.  At
$z=8$ this could in principle reflect the presence of sightlines that 
subtend neutral regions, but by $z=5$ it indicates the impact of voids 
on absorber statistics.  That the distributions are not significantly 
more skewed to low numbers at $z=8$ than at $z=5$ indicates that voids 
play a bigger role in generating the skew than neutral regions because 
carbon is ionized locally even in regions that are neutral on large 
scales.

\subsection{Evolution in Gas or UVB?}\label{ssec:whatEvolves}
\begin{figure}
\centerline{
\setlength{\epsfxsize}{0.5\textwidth}
\centerline{\epsfbox{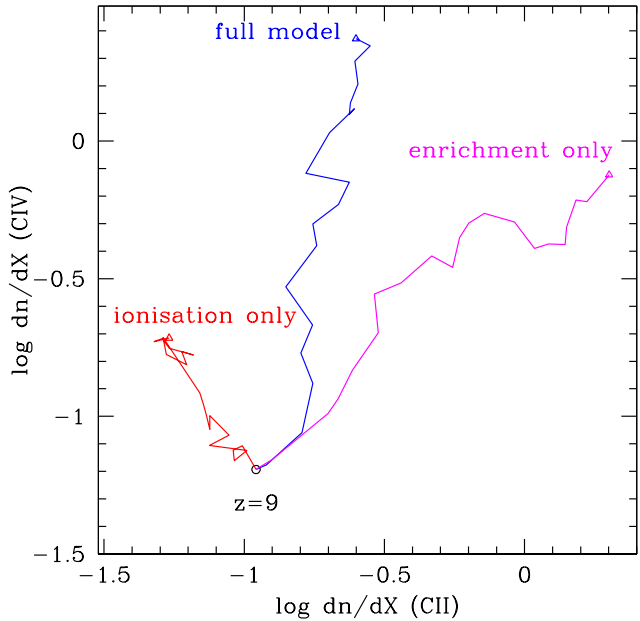}}
}
\caption{The predicted evolution in the number of absorption systems per path
length from $z=9\rightarrow5.5$ in our fiducial case (blue) versus a toy model 
in which no extra metals are ejected after $z=9$ (red) and a model where the 
UVB does not evolve after $z=9$ (magenta).  Evolution in the UVB and the CGM 
both contribute to the overall evolution in absorber abundance.
}
\label{fig:whatEvolves}
\end{figure}

A central question in interpreting metal absorbers regards what drives their
evolution.  The three major contributing factors are the growing UVB, the 
declining mean density of CGM gas, and ongoing enrichment.  If evolution is 
primarily driven by a growing UVB, then (1) metal absorbers are a 
complementary tracer of reionization~\citep{oh02,fur03}; and (2) the 
observed metals are the signature of a very early generation of stars.  
If evolution is primarily driven by a tradeoff between continuing enrichment
and declining densities, then observations trace ongoing star formation in 
low-mass galaxies~\citep{bec11}. 


In order to address this question, we re-compute the predicted number of $\cii$ 
and $\civ$ absorbers per path length in the column density range of 
$10^{13}$--$10^{15}$cm$^{-2}$ ($\lcii$ and $\lciv$, respectively) in two 
toy-model scenarios.  First, we take the gas density and metallicity from our 
$z=9$ snapshot while allowing the UVB to evolve down to $z=5.5$.  This 
``ionization-only" scenario is illustrated by the red trajectory in 
Figure~\ref{fig:whatEvolves} leading upwards and to the left.  
Quite intuitively, a growing UVB suppresses $\lcii$ and 
boosts $\lciv$.  In the second toy-model, we allow the gas density and metallicity 
to evolve but leave the UVB fixed (in proper units) to the $z=9$ field.  This 
``enrichment-only" model is indicated by the magenta trajectory leading to 
increasing $\lcii$ and $\lciv$.  In this case, there is fairly continuous
growth in both $\lcii$ and $\lciv$.  The full model runs between these two.
We conclude that the evolving population of carbon absorbers does not 
predominantly trace either UVB or CGM evolution because they either cancel 
(in the case of $\lcii$) or both drive signicant evolution (in the case 
of $\lciv$).  Instead, it is sensitive to both.

\subsection{Large-Scale Fluctuations in the UVB}\label{ssec:UVBfluct}
\begin{figure}
\centerline{
\setlength{\epsfxsize}{0.5\textwidth}
\centerline{\epsfbox{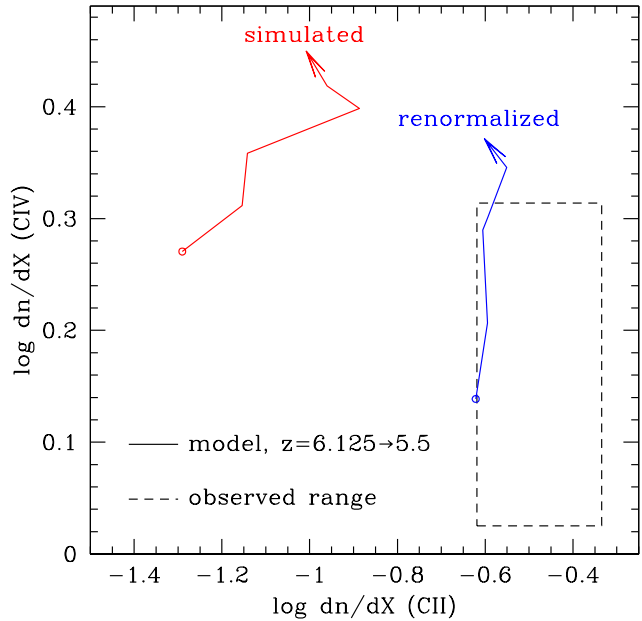}}
}
\caption{The impact of our renormalized field on the predicted number of
systems per path length from $z=6\rightarrow5.5$.  Renormalizing the UVB to 
boost $\tau_{\mathrm{Ly}\alpha}$ suppresses $\civ$ modestly while boosting 
the number of $\cii$ systems by $\approx4\times$.  The dashed box shows the 
observationally-allowed region from Figure~\ref{fig:dndX_predict}.
}
\label{fig:Juncert}
\end{figure}

The observable space in Figure~\ref{fig:whatEvolves} is a convenient
setting for exploring how large-scale spatial fluctuations in the UVB
might impact our predictions.  In particular,~\citet{bec14} have
recently argued that the UVB may be inhomogeneous down to $z=5$ on scales
that are not captured by our simulation.  We showed in Figure~\ref{fig:tau_v_z}
that the simulated and renormalized UVBs roughly span the observed range
of $\tau_{\mathrm{Ly}\alpha}$ measurements from $z=5$--6.  We now use this 
fact to bracket the uncertainty owing to large-scale UVB fluctuations.  
In Figure~\ref{fig:Juncert}, we show how $\lcii$ and $\lciv$
evolve from $z=6.125\rightarrow5.5$ using the simulated (left) and renormalized 
(right) UVBs, while the dashed box indicates the same observations as in
Figure~\ref{fig:dndX_predict}.  The simulated UVB overproduces $\lciv$ and 
underproduces $\lcii$, indicating that its amplitude is on average too high.  
Renormalizing the UVB increases $\lcii$ by a factor of 3--4 and reduces 
$\lciv$ abundance by 0.1 dex.  To the 
extent that the simulated and renormalized UVBs span the realistic range, the
converged value will lie somewhere between these curves.  On the other hand, 
given that the renormalized field is not weak enough to match the regions 
where flux is undetected in Figure~\ref{fig:tau_v_z}, the true correction 
may be somewhat larger still.

\section{Summary}\label{sec:sum}
\subsection{Summary of Results}\label{ssec:res}
We have used a cosmological hydrodynamic simulation that models a 
spatially-inhomogeneous
multifrequency ionizing background on-the-fly to study how the
abundance of $\cii$ and $\civ$ absorbers evolves through the latter 
half of the hydrogen reionization epoch.  Our results are as follows:
\begin{itemize}
\item The volume average of our simulated radiation field is within a 
factor of 2--4$\times$ of the spatially-uniform HM12 model 
down to $z=3$; this discrepancy is not large compared to observational
uncertainties.
\item Consistently with observations, the predicted UVB at the HI 
ionization edge does not evolve strongly between $z=$6--3.
\item The mean radiation field shows large spatial fluctuations on
scales that metal absorbers are sensitive to and is stronger than the
volume-average in regions where metals lurk,
indicating that models cannot predict the abundance of metal absorbers
to better than 20--30\% accuracy unless they account for local sources.
\item Our model matches the observed $\civ$ and $\cii$ abundances at 
$z=6$ and slightly overproduces low-column  $\civ$ absorbers at $z=3$.
Meanwhile, it does not reproduce the observed mass 
densities at $z\sim6$, indicating that further work is necessary to
constrain the CDD slopes. 
\item The volume-averaged mass fraction, absorber abundance, and mass
density of $\civ$ decrease to high redshifts.  Meanwhile, the $\cii$ 
mass fraction increases owing to weakening UVB and the increasing 
proper density of metal-bearing regions.  However, the overall $\cii$ 
absorber abundance and ion mass density decrease slowly with increasing 
redshift because the CGM metallicity decreases.  
\item The abundances and mass densities of $\cii$ and $\civ$ are 
predicted to cross at $z=8$,
reflecting evolution in the volume-averaged mass fractions.  
\item The predicted number of low-ionization absorbers per quasar sightline shows 
roughly the same scatter as observations, with the implication that the
observed scatter does not require a partially-neutral universe at
$z=6$.  There are somewhat more sightlines with zero absorbers than 
predicted by Poisson statistics, indicating the influence of voids.
\end{itemize}

\subsection{Summary of Limitations}\label{ssec:lim}
The major drawback of our simulation is its small volume, which
introduces three limitations.  First, we saw in Section~\ref{sec:J} that our 
simulation predicts an artificially rapid rise in the UVB amplitude at 
the overlap epoch.  We compensate for 
this by normalizing the simulated UVB so that its volume average does not
exceed the HM12 model at the $\hi$ ionization edge.  In practice, 
this means scaling the entire simulated UVB at each redshift below $z=6.5$
by a factor that ranges from 2 to 4.  This calibration preserves the 
simulated UVB's spectral slope and spatial fluctuations while preventing the 
artefact from propagating into our predictions.  

Unfortunately, Figure~\ref{fig:tau_v_z} shows that this calibration
does not bring the predicted Lyman-$\alpha$ optical depth into complete 
agreement with observations owing to the second problem, which is an
inaccurate sampling of voids and overdensities.  
Lyman-$\alpha$ transmission occurs in voids where the neutral fraction 
is lower~\citep{bol09}, hence it requires large volumes.  Meanwhile, 
metal absorbers occur in overdensities where galaxies form, hence they
require high resolution in order to capture the dominant, low-mass
galaxy population.

The final consequence of our small volume is the incomplete overlap 
between the simulated and observed column density ranges at $z=6$.  This
occurs  because our simulation systematically lacks the massive halos that 
host high-column systems (Figures~\ref{fig:dNdX_CII}--\ref{fig:dNdX_CIV}).  
We compensate for this by extrapolating the predicted CDDs assuming 
power-law slopes of -1.7. 

In short, increasing our dynamic range will simultaneously yield a smoother
reionization history, a more realistic representation of voids and
overdensities, and a more accurate model for the CDD.  These 
improvements will enhance our ability to understand how metal absorbers
probe the galaxies, enriched CGM, and UVB that had formed by the close of the 
reionization epoch.

\section*{Acknowledgements}
We thank J.\ Hennawi, M.\ Prescott, and R. Cen for helpful conversations.  
We are indebted to V.\ Springel for making {\sc Gadget-3}
available to our group, and for P.\ Hopkins for kindly sharing his SPH 
module.  We thank for the anonymous referee for a thoughtful 
report that improved the draft.  Our simulation was run on the {\sc Steno}
facility at the University of Copenhagen, and we are indebted to the support
staff at hpc{@}ucph for their support.  KF thanks the Danish National
Research Foundation for funding the Dark Cosmology Centre.  EZ acknowledges 
research funding from the Swedish Research Council, the Wenner-Gren 
Foundations and the Swedish National Space Board.

\setcounter{subsection}{0}
\appendix
\section*{Appendix}
\renewcommand{\thesubsection}{\Alph{subsection}}
\subsection{Test of self-shielding}\label{app:ss}

\begin{figure}
\centerline{
\setlength{\epsfxsize}{0.5\textwidth}
\centerline{\epsfbox{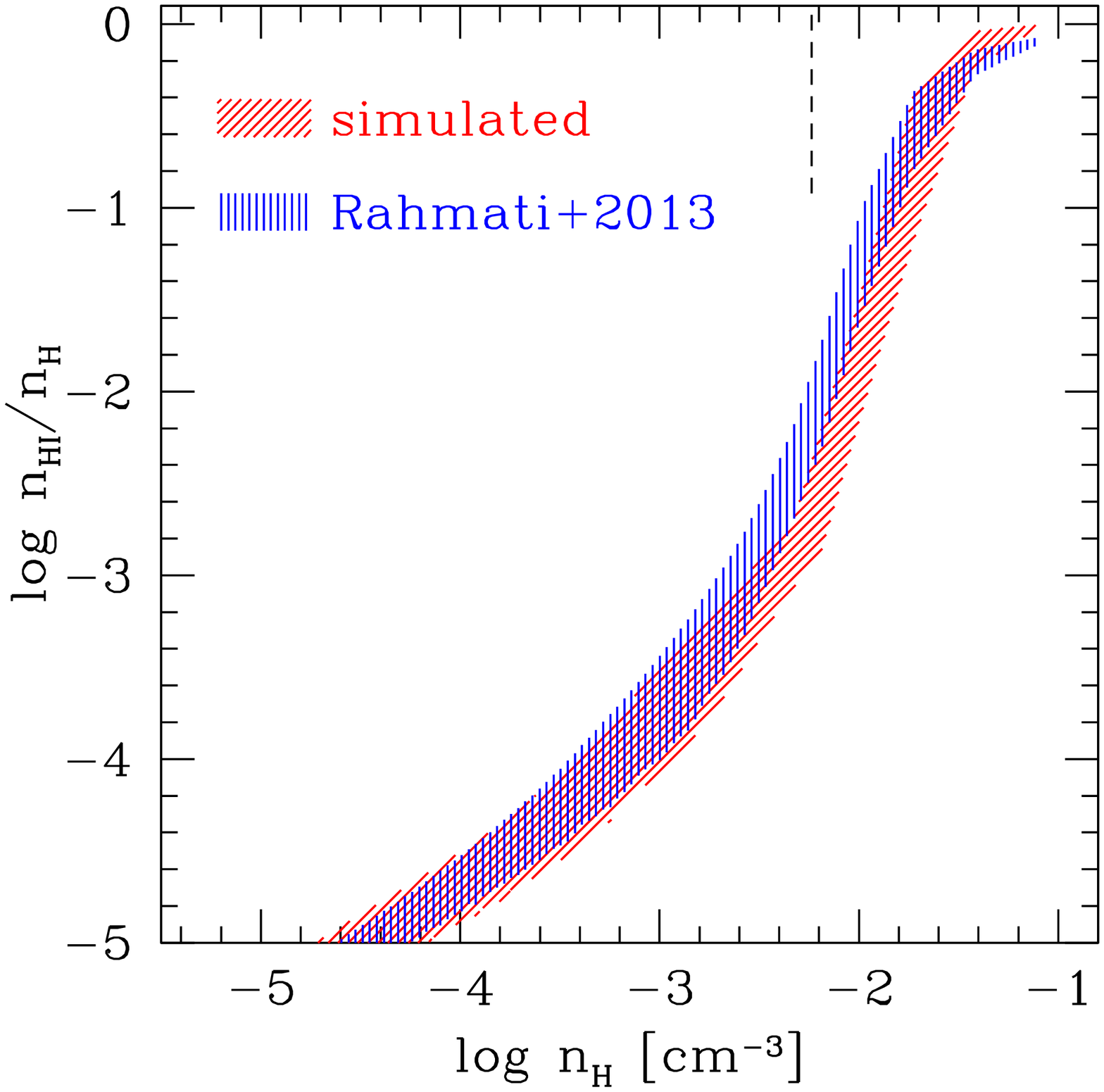}}
}
\caption{The 15--85\% neutral hydrogen fraction range versus proper hydrogen 
number density at $z=3$ in our simulation (red, diagonal shading) and using 
the~\citet{rah13} attenuation function (blue, vertical shading).  The vertical 
dashed segment indicates a representative self-shielding threshold density of 
0.006 cm$^{-3}$ (assuming a temperature of $10^4 K$).  The two ranges overlap 
at all densities, indicating that our self-shielding treatment attenuates the 
radiation field realistically.
}
\label{fig:xHIvnH}
\end{figure}

As a test of our self-shielding prescription, we compare the simulated 
trend of neutral hydrogen fraction versus proper hydrogen number density 
$n_H$ at $z=3$ with the spatially-resolved study 
by~\citet{rah13}.  We implement the~\citet{rah13} attenuation model as 
follows: First, we compute the volume-averaged mean hydrogen 
photoionization rate directly from our simulation (their Equation 3).  
Next, we use their Equation 13 to compute the threshold density for 
self-shielding, adopting each particle's local temperature in turn.  Next, 
we use their Equation 14 to compute the attenuation.  Finally, we use the
attenuated photoionization rate to compute the equilibrium ionization 
fraction following~\citet{kat96}.  
Note that we do not employ the renormalised radiation field 
for this purpose (Section~\ref{sec:J}), hence the comparison is a direct 
test of the simulation.

Figure~\ref{fig:xHIvnH} shows that the two models are in broad agreement, 
indicating that our self-shielding model is realistic.   In detail,
however, the simulated neutral fraction is up to $3\times$ 
lower for densities $-2.5 < \log(n_H) < -1.5$, while for higher 
densities it is somewhat higher.  We have directly verified that
these slight differences do not reflect the helium ionization state, 
the predicted flux above 4 Ryd, or departures from ionization 
equilibrium.  Instead, they suggest (1) that the simulated radiation 
field is insufficiently attenuated at densities near the self-shielding 
threshold; and (2) that very dense gas ($-1.5 < \log(n_H)$) is slightly 
too neutral in the simulation because we omit ionizing recombination 
radiation (see, for example, Figure 4 of~\citealt{rah13}).  The lack of 
recombination radiation is not a problem for our study because it is
sharply peaked at 1 Ryd whereas the $\cii$ photoionization threshold
is 1.8 Ryd.  Hence while there is room for progress in accounting
for small-scale radiative transfer effects accurately, we conclude that 
our simulation is adequate for our present purposes.

\end{document}